\documentclass[a4paper,aps,pra,twocolumn,showpacs,superscriptaddress]{revtex4}
\usepackage{amsmath,amssymb,latexsym}
\usepackage[english]{babel}
\usepackage{graphicx}
\newtheorem{lemma}{Lemma}
\newtheorem{theorem}{Theorem}
\newtheorem{corollary}{Corollary}
\newtheorem{fact}{Fact}
\newcommand{\tr}{\textrm{Tr}}

\newcommand{\la}{\lambda}
\newcommand{\id}{\mathbb I}

\usepackage{geometry}
\begin{document}

\title{Multipartite entanglement detection from correlation tensors}
\author{Julio I. de Vicente}  \email{julio.de-vicente@uibk.ac.at}
\affiliation{Institut f\"ur Theoretische Physik, Universit\"at Innsbruck, Technikerstr.\ 25, 6020 Innsbruck, Austria}
\author{Marcus Huber}  \email{marcus.huber@univie.ac.at}
\affiliation{Faculty of Physics, University of Vienna, Boltzmanngasse 5, 1090 Vienna, Austria}

\begin{abstract}
We introduce a general framework for detecting genuine multipartite entanglement and non full-separability in multipartite quantum systems of arbitrary dimensions based on correlation tensors. Regarding genuine multipartite entanglement our conditions are comparable to previous approaches in the case of qubits while they show particular strength in the relatively unexplored case of higher dimensional systems. In the case of non-full separability our conditions prove to be advantageous in situations where more than two-body correlations are relevant, where most previous conditions turned out to be weak. Moreover, they allow for the detection of fully bound entangled states. Finally, we also discuss experimentally-friendly ways of implementing our conditions, which are based on directly measurable quantities.
\end{abstract}


\maketitle
\section{Introduction}

In many-body quantum physics entanglement constitutes a fundamental feature. Complex systems with multipartite quantum correlations can be exploited to enable numerous tasks in quantum
information processing. The multipartite entanglement in these systems enables quantum computation (e.\ g.\ \cite{qc}), multi-party cryptography (e.\ g.\ \cite{HBB,SHH1,gisin-crypt}) and the implementation of various other quantum algorithms (e.\ g.\ \cite{brussqa}). Apart from these possible applications in modern quantum technologies it has become apparent that multipartite entanglement also plays a fundamental role in the physics of complex systems. While the involvement in quantum phase transitions (e.\ g.\ \cite{phase}) and ionization procedures (e.\ g.\ \cite{helium}) seems clear, the recently suggested role in biological systems is still subject of debate (e.\ g.\ \cite{Caruso,bio,DNA,NotBio}).\\

Therefore, to decide if a state is entangled or not is a fundamental problem in quantum information theory \cite{horodeckiqe,guehnewit}. Though a simple mathematical characterization is elusive (the problem has in fact been proved to be NP hard \cite{Gur}), several works have put up sufficient conditions to identify a multipartite state as entangled \cite{horodeckicrit,mbloch,ossi,git}. These conditions are not only helpful for entanglement detection but also they provide more physical insight into this phenomenon. Contrary to the bipartite case, there exist different classes of multipartite entangled states. Genuine multipartite entanglement is of particular interest since it involves entanglement between \textit{all} the subsystems. Recently there has been a lot of progress concerning its detection, mostly using linear and non-linear entanglement witnesses \cite{linwit,seevinckcrit,strfac,guehnecrit,HMGH1,GHH1,HSGSBH1,HESGH1,Guehnetaming} and Bell-like inequalities \cite{belllike,diew}. However, with a few exceptions (see e.\ g.\ \cite{seevinckcrit}), the approaches taken in each particular case only allow to detect either entanglement or genuine multipartite entanglement. Moreover, most of them are limited to qubit systems.

In this paper we develop a general framework which allows using the same piece of information to detect both entanglement and genuinely multipartite entanglement for multipartite states of arbitrary dimensions. Our main tool will be correlation tensors which are built from the expectation values of a local operator basis. Our motivation stems from different facts. First, it has been shown that all information about the entanglement properties of a system is encoded in the correlation tensors \cite{paterek} and these mathematical tools have already been proven useful for the detection of entanglement in the bipartite case \cite{bloch}. In \cite{mbloch} a first step towards the extension of these ideas to the detection of multipartite entanglement has been taken (see also \cite{john} for a correlation-tensor approach to multipartite entanglement detection). However, here we will show that this allows for a much more general formalism (in which the criterion of \cite{mbloch} is a particular case and that of \cite{john} is strictly weaker), which, furthermore, enables to identify different classes of multipartite entanglement. Our conditions are expressed through simple mathematical inequalities. Contrary to entanglement witnesses, which are designed for a particular class of states, violations of these inequalities signal genuine multipartite entanglement or non-full separability for general states. Moreover, since the entries of the correlation tensors are directly related to measurable quantities, we will discuss how our approach can be adapted to optimize the experimental effort. Last, many conditions for multipartite entanglement such as spin squeezing inequalities \cite{ossi}, covariance matrices \cite{git}, entanglement witnesses based on structure factors \cite{strfac} or two-particle Hamiltonians \cite{toth} or magnetic susceptibility measurements \cite{wies} rely only on two-body correlations. It has been shown in \cite{git} that this limits their ability to detect entanglement as there are important classes of states like graph states which have the same two-particle reduced states as separable states. Hence, their entanglement cannot be reveled by just looking at two-point correlations. On the other hand, correlation tensors take into account all $m$-body correlations. This suggests (and we will later see) that correlation tensors may overcome the limitations of the previous criteria.

  %If one seeks to estimate the relevance of multipartite entanglement, one must go beyond mere detection and also quantify the amount of multipartite entanglement contained in the system. Recent attempts to achieve such a quantification were proposed using entanglement witnesses and optimization in Refs.~\cite{crazychin,Guehnetaming}. However, especially for very large dimensional systems, such an optimization is computationally unfeasible and thus very limited. In order to have a reliable criterion that also allows quantification it is important to develop criteria which do not need to rely on any optimization.\\
%Exactly this is the goal of this article. As it was shown that norms of the correlation tensor provide strong criteria in bipartite quantum systems (see e.g. Refs.~\cite{bloch,mbloch,yojpa}), we want to extend these concepts to derive criteria detecting and quantifying both bi- and multipartite entanglement in high dimensional systems.\\

\section{Preliminaries}
Before we proceed to derive our main results let us briefly review the definitions of multipartite entanglement and correlation tensors. We consider $n$-partite quantum states $\rho$ acting on the Hilbert space $H=H_1\otimes\cdots\otimes H_n$ of dimension $D=d_1\cdots d_n$. If a pure state $|\Psi\rangle\in H$ can be written as a tensor product of states for \textit{every} subsystem, i.\ e.\
\begin{equation}
|\Psi\rangle\langle \Psi|=|\psi_{1}\rangle\langle\psi_{1}|\otimes\cdots\otimes|\psi_{n}\rangle\langle\psi_{n}|,
\end{equation}
then the state is said to be fully separable. Consequently, fully separable mixed states are convex combinations of fully separable pure states. These states contain no entanglement \textit{at all}. On the other hand, any $n$-partite pure state that can be written as a tensor product
\begin{equation}
|\Psi\rangle\langle \Psi|=|\Psi_{A}\rangle\langle\Psi_{A}|\otimes|\Psi_{\bar{A}}\rangle\langle\Psi_{\bar{A}}|
\end{equation}
with respect to some bipartition $A\bar{A}$ ($A$ denoting some subset of subsystems and $\bar{A}$ its complement) is called biseparable. These states might contain some entanglement (as $|\Psi_{A}\rangle$ and/or $|\Psi_{\bar{A}}\rangle$ might not be separable) but they are not completely entangled. States that are not biseparable with respect to any partition are then said to be genuinely multipartite entangled. The generalization to mixed states is straightforward. Any mixed state that can be decomposed into a convex sum of biseparable pure states is called biseparable. Consequently, any non-biseparable mixed state is called genuinely multipartite entangled. Due to the fact that the bipartitions might differ for every element of the biseparable decomposition it is an intricate task to find out whether such a decomposition is possible.\\
Let $\{\la_i^{(j)}\}_{i=1}^{d^2_j-1}$ denote the generators of $SU(d_j)$ and let $\la_0^{(j)}=\id_{d_j}$, which altogether constitute an orthogonal basis of the real Hilbert-Schmidt space of Hermitian operators acting on $H_j$ (i.\ e.\ with inner product $\langle A,B\rangle=\tr(AB)$). Thus, so is $\{\bigotimes_{j=1}^n\{\la_i^{(j)}\}\}$ for the operators acting on $H$ and, hence, $\rho$ is completely characterized by the expectation values $\langle\lambda^{(1)}_{i_1}\otimes\cdots\otimes\la^{(n)}_{i_n}\rangle:=\mathcal{T}_{i_1\cdots i_n}$ where $i_j=0,1,\ldots,d_j^2-1$, which gives rise to the so-called (multipartite) Bloch representation or (multipartite) Fano form of density operators \footnote{Notice that $\mathcal{T}_{0\cdots0}$ is fixed by the normalization condition $\tr\rho=1$ and there are indeed $\prod_jd_j^2-1$ parameters.}. We will decompose the tensor $\mathcal{T}_{i_1\cdots i_n}$ into the $m$-body correlation tensors $T^{(j)}_{i_j}$, $T^{(j,k)}_{i_ji_k}$, etc, which are tensors of order $m$ indicated by the number of labels in the superscript. All the indices not labeled in the superscript are fixed to be zero while the other indices take every possible value but zero (i.\ e.\ the identity is not taken into account). For instance, the 1-body correlation tensor for particle 1, given by $T^{(1)}_{i_1}=\mathcal{T}_{i_10\cdots0}$ with $i_1\neq0$, completely characterizes the reduced state $\rho_1$ and the 2-body correlation tensor for subsystems 1 and 2, $T^{(1,2)}_{i_1i_2}=\mathcal{T}_{i_1i_20\cdots0}$ ($i_1,i_2\neq0$), together with the 1-body correlation tensors of 1 and 2 characterizes $\rho_{12}$ and so on. For the $n$-body correlation tensor, which we shall also call full correlation tensor, we will drop the superscripts to ease the notation, i.\ e.\ $T_{i_1\cdots i_n}=\mathcal{T}_{i_1\cdots i_n}$ ($i_j\neq0$ $\forall\,j$).

Given two tensors $\mathcal{T}_{i_1\cdots i_n}$ and $\mathcal{S}_{j_1\cdots j_m}$, their outer product $\circ$ is the $(n+m)$th order tensor $(\mathcal{T}\circ \mathcal{S})_{i_1\cdots i_nj_1\cdots j_m}=\mathcal{T}_{i_1\cdots i_n}\mathcal{S}_{j_1\cdots j_m}$. If some tensor can be written as the outer product of two other tensors, say $\mathcal{T}_{i_1\cdots i_n}=\mathcal{R}_{i_1i_2i_3}\mathcal{S}_{i_4\cdots i_n}$, we will say that the tensor factorizes in the corresponding splitting ($\{1,2,3\},\{4,5,\ldots n\}$ in this case). If a tensor cannot be written as the outer product of any two lower order tensors we will say that the tensor does not factorize.

It has been shown in \cite{bloch} that a bipartite pure state is separable if and only if (iff) $T^{(1,2)}_{i_1i_2}=T^{(1)}_{i_1}T^{(2)}_{i_2}$. Accordingly, a multipartite pure state is biseparable with respect to the partition $A\bar{A}$ iff $\mathcal{T}^{(A,\bar{A})}_{i_Ai_{\bar{A}}}=\mathcal{T}^{(A)}_{i_A}\mathcal{T}^{(\bar{A})}_{i_{\bar{A}}}$. Thus, we have the following characterization of biseparable pure states:
\begin{fact}
A pure state is biseparable iff there exists some partition of the subsystems $A\bar{A}$ for which all the $m$-body correlation tensors involving $k$ particles from $A$ and $m-k$ from $\bar{A}$ ($k\neq0,m$) factorize into the corresponding $k$-body correlation tensor of the $k$ particles from $A$ and the $(m-k)$-body correlation tensor of the $m-k$ particles from $\bar{A}$.
\end{fact}

This leads to a simple sufficient condition for genuinely multipartite entangled pure states:
\begin{corollary}
If some $m$-body correlation tensor of a pure state cannot be factorized into meaningful lower order correlation tensors, then the state contains genuine multipartite entanglement.
\end{corollary}

Analogously, this can be extended to non fully separable states (see also \cite{mbloch}):
\begin{corollary}
If some $m$-body correlation tensor of a pure state cannot be fully factorized into meaningful 1-body correlation tensors, then the state is not fully separable.
\end{corollary}

We stress that the factorization must be possible into meaningful correlation tensors. This a consequence of the fact that the Bloch representation holds for Hermitian operators and not only for density operators, which are furthermore positive semidefinite. Hence, not all values of $\mathcal{T}_{i_1\cdots i_n}$ give rise to a density matrix, i.\ e.\ are meaningful. To characterize this subset is a quite involved problem (see e.\ g.\ \cite{blochchar}). However, there exist several conditions the set of meaningful correlations should fulfill. For instance, it will be useful later on that for $1$-body correlation tensors it must hold that
\begin{equation}\label{snorm1corr}
||T^{(j)}||\leq\sqrt{\frac{2(d_j-1)}{d_j}},
\end{equation}
with equality iff the state is pure and where $||\cdot||$ is the standard Euclidean norm for vectors. This expresses the fact that $\tr\rho_j^2\leq1$. This condition holds for the following choice of normalization for the generators of $SU(d_j)$: $\tr(\la_m\la_n)=2\delta_{mn}$ (of course $m,n\neq0$ since for the identity we have $\tr(\la^{(j)}_0\la^{(j)}_0)=d_j$). We will follow this convention throughout the paper, with which for qubits the generators correspond to the standard Pauli matrices.

Following \cite{bloch,mbloch}, the main idea behind this paper is to express the factorizability of some tensor into lower order meaningful tensors as an upper bound on some convex function. Convexity will then imply that this bound must hold as well for biseparable (fully separable) mixed states and, hence, a violation of this bound will signal the presence of genuine multipartite entanglement (non full separability) for general quantum states. It seems that some tensor norm is the best choice of convex function since the norm of meaningful correlation tensors is upper bounded as we have just seen for 1-body correlation tensors. Notice that convexity in this case is guaranteed by the triangle inequality. Physical intuition suggests that full correlation tensors should be the first ones to check and usually we will restrict ourselves to them.

The rest of the paper is organized as follows. In Secs. \ref{standard} and \ref{matr} we provide two different approaches that lead to conditions for the identification of genuine multipartite entanglement. In Sec.\ \ref{full} we show that similar techniques can be used to obtain conditions for non full separability. Section \ref{props} is devoted to some mathematical properties of our conditions which are related to their experimental implementation. Final conclusions are drawn in Sec.\ \ref{conclusions}.

\section{Genuine multipartite entanglement conditions based on the standard tensor norm}\label{standard}

The standard tensor norm is defined as the natural generalization of the Euclidean vector norm to higher order tensors (recall that we will always deal with real tensors), i.\ e.\
\begin{equation}\label{snorm}
||T_{i_1\cdots i_n}||^2=\sum_{i_1,\ldots,i_n}T^2_{i_1\cdots i_n}.
\end{equation}
This seems to be a very good choice for our purposes since this norm is multiplicative under outer products, i.\ e.\ $||T\circ S||=||T|| ||S||$ $\forall\,T,S$. Hence, we will just need to upper bound the standard norm of the $m$-body correlation tensors. This turns out to be quite easy. As we mentioned above, the condition that $\tr\rho_j^2\leq1$ must hold translates into an upper bound for the standard norm of the 1-body correlation tensors. Now, combining this conditions with $\tr\rho_{ij}^2\leq1$ will yield an upper bound for the 2-body correlation tensors (see e.\ g.\ \cite{yojpa}). This procedure can be recursively applied to upper bound the standard norm of all meaningful $m$-body correlation tensors. For instance, this gives
\begin{equation}\label{snorm2corr}
||T^{(j,k)}||\leq2\sqrt{\frac{d_jd_k-1}{d_jd_k}},
\end{equation}
with equality iff the state $\rho_{jk}$ is a maximally entangled state \cite{yojpa}.

To illustrate this, let us start by considering a tripartite pure state with subsystems of equal dimension $d_j=d$ $\forall\,j$. Then, full separability implies
\begin{equation}
||T_{i_1i_2i_3}||=||T^{(1)}||||T^{(2)}||||T^{(3)}||=\left(\frac{2(d-1)}{d}\right)^{3/2},
\end{equation}
and biseparability between any two subsystems and the other yields
\begin{equation}
||T_{i_1i_2i_3}||=||T^{(j)}||||T^{(k,l)}||\leq\sqrt{\frac{8(d-1)(d^2-1)}{d^3}}.
\end{equation}
Since the last condition is more restrictive and using convexity we then have
\begin{theorem}\label{thsnorm} If for an arbitrary (pure or mixed) tripartite state it holds that
\begin{equation}\label{snormcond}
||T_{i_1i_2i_3}||>\sqrt{\frac{8(d-1)(d^2-1)}{d^3}},
\end{equation}
then the state is genuinely multipartite entangled.
\end{theorem}

\begin{figure*}[t!]
\begin{center}
(a)
\includegraphics[width=5cm, keepaspectratio=true]{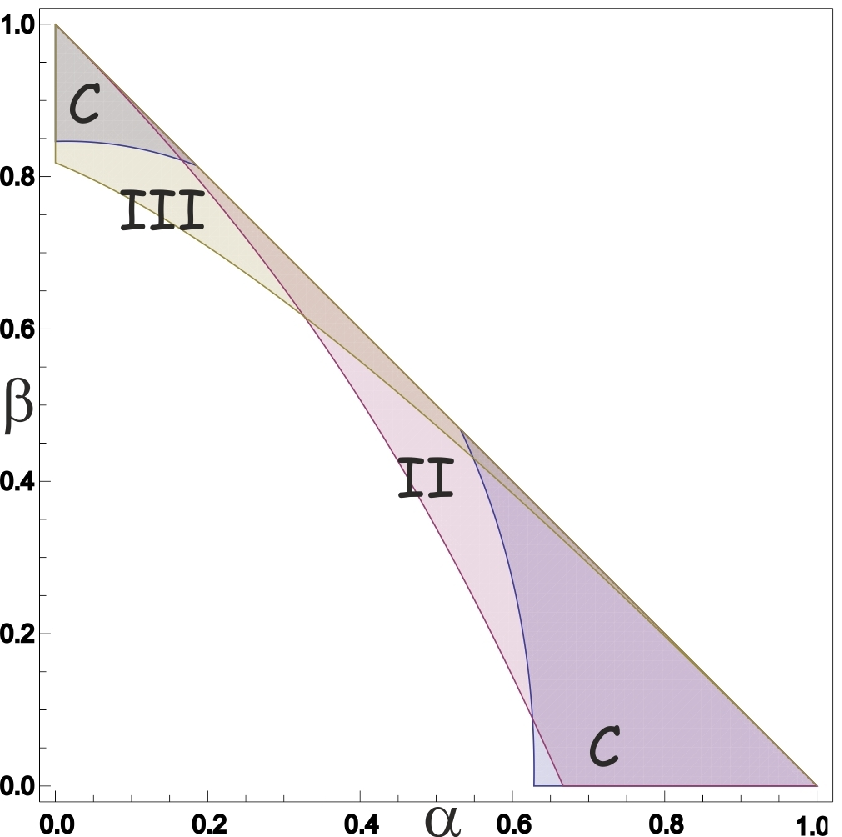}
(b)
\includegraphics[width=5cm, keepaspectratio=true]{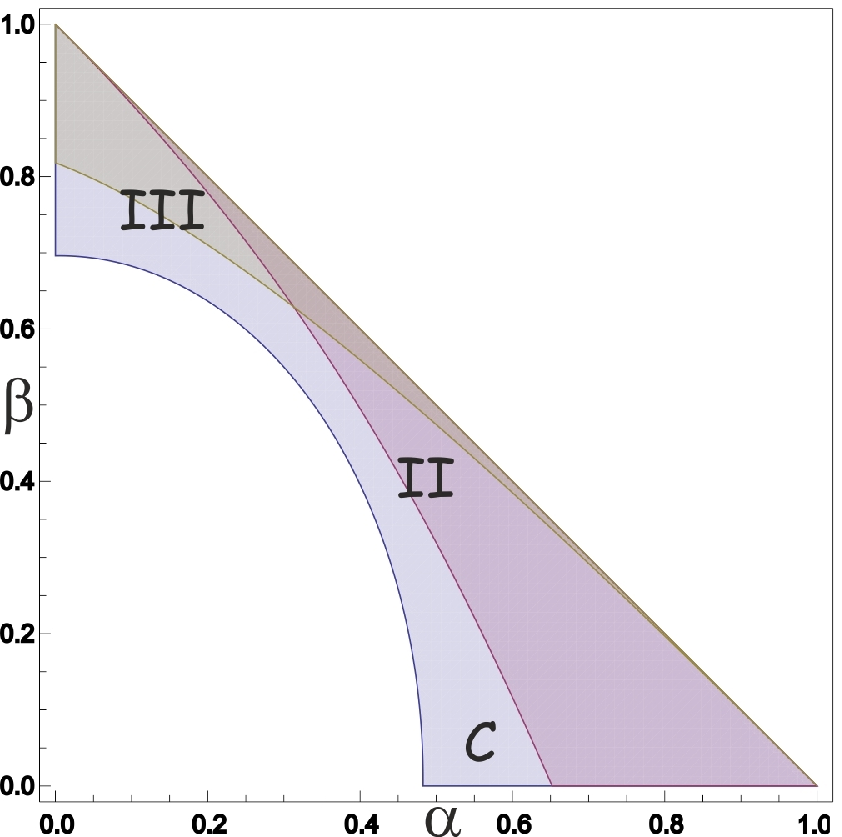}
(c)
\includegraphics[width=5cm, keepaspectratio=true]{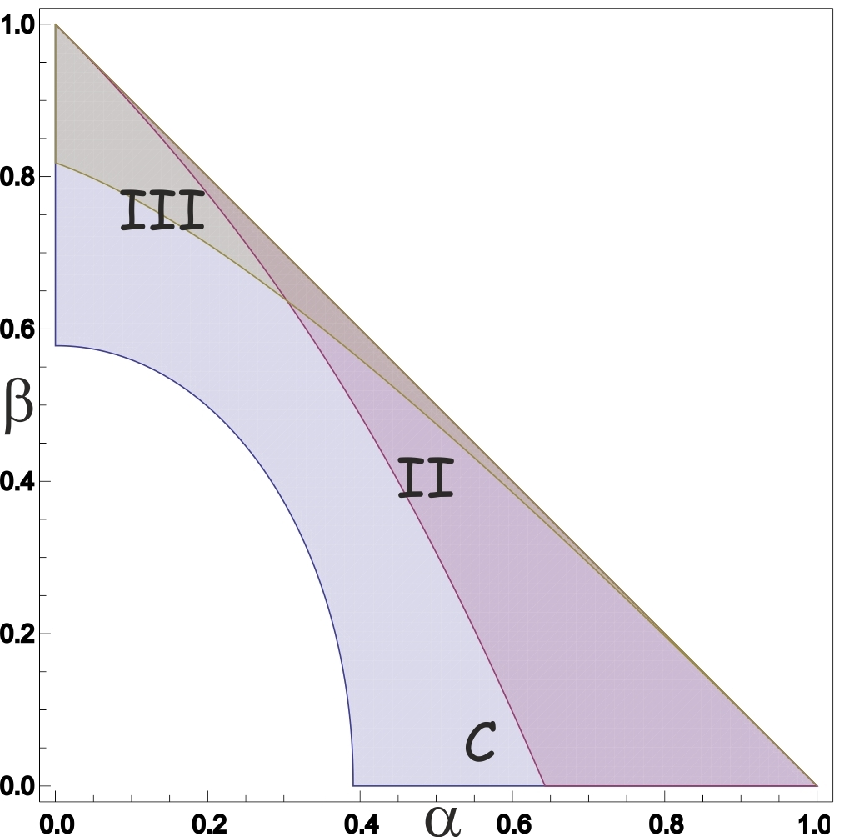}
\caption{(Color online) Here the parameter regions for which the state $\rho=\alpha \rho_{GHZ(d)}+\beta \rho_{W(d)}+\frac{1-\alpha-\beta}{2d-2}(\sum_{i=0}^{d-1}|i,i,i+1\rangle\langle i,i,i+1|+|i+1,i+1,i\rangle\langle i+1,i+1,i|)$ for (a)$d=4$, (b)$d=5$ and (c)$d=6$ exhibits genuine multipartite entanglement are identified. The generalized GHZ and W states for $d$-dimensional systems are defined as $\rho_{GHZ(d)}=|GHZ(d)\rangle\langle GHZ(d)|$ with $|GHZ(d)\rangle:=\frac{1}{\sqrt{d}}\sum_{i=0}^{d-1}|i,i,i\rangle$ and $\rho_{W(d)}=|W(d)\rangle\langle W(d)|$ with $|W(d)\rangle=\frac{1}{\sqrt{3(d-1)}}\sum_{i=0}^{d-2}(|i,i,i+1\rangle+|i,i+1,i\rangle+|i+1,i,i\rangle)$. The (red) region labeled II, uses criterion II from Ref.~\cite{HMGH1} optimized numerically over all local unitary representations of the density matrix. The (yellow) region labeled III, uses criterion III from Ref.~\cite{HMGH1} optimized numerically over all local unitary representations of the density matrix. The numerical optimization was performed using the composite parametrization from Ref.~\cite{compp}. The (blue) region labeled C, shows the states detected to be genuinely multipartite entangled using Eq.\ (\ref{snormcond}).}\label{example2}
\end{center}
\end{figure*}

This simple mathematical idea is already strong enough to detect paradigmatic cases of genuine multipartite entanglement. If we consider (\ref{snormcond}) for three qubits, this gives the
bound $\sqrt{3}\simeq1.73$ while the GHZ and W states have respectively $||T_{i_1i_2i_3}||=2$ and $||T_{i_1i_2i_3}||\simeq1.92$. Therefore, their genuine multipartite entanglement is
successfully identified, leading to (modest) white noise tolerances of $p_{GHZ}\lesssim0.13$ and $p_W\lesssim0.10$ \footnote{Here and throughout the paper the white noise tolerance of some
entanglement condition for a state $\psi$ is defined as the values of $p$ for which $p\,\mathbb{I}_D/D+(1-p)|\psi\rangle\langle\psi|$ is still detected by this condition.}. Furthermore, the
power of this condition increases with the subsystem dimension improving remarkably on \cite{HMGH1}. In Figure \ref{example2} we plot the detection power of Eq.\ (\ref{snormcond}) for dimensions
$4$, $5$ and $6$. As the other criteria for genuine multipartite entanglement in high dimensional systems are still based on qubit subsystems of the high dimensional Hilbert space it is perhaps
not surprising that our criteria, exploiting all degrees of freedom, quickly outperform them with growing dimensionality of the system.

As discussed above the extension of this condition to states with more subsystems or different subsystem dimensions is straightforward. However, for four qubits we have for both the GHZ and the Dicke state with two excitations $||T_{i_1i_2i_3i_4}||=3$, which is precisely the same value of a tensor product of two maximally entangled bipartite states ($||T_{i_1i_2i_3i_4}||=\sqrt{3}\cdot\sqrt{3}$). On the analogy of (\ref{snormcond}), one might hope to improve for larger $d$; nevertheless, in the next section we will present a different and more powerful approach.

It is worth mentioning that $||T_{i_1\cdots i_n}||-[2(d-1)/d]^{n/2}$ has been shown to be an entanglement monotone \cite{blochmes}. Our results show that a high value of this measure can not only imply some entanglement but even genuine multipartite entanglement.

\section{Genuine multipartite entanglement conditions based on norms of matricizations of tensors}\label{matr}

As an alternative to the previous section one can seek for other norms. Unfortunately, to our knowledge, the standard norm is the only norm which is multiplicative under outer products, a property which is very convenient for the mathematical simplicity of our derivations. Nevertheless, it turns out that considering matricizations of tensors (i.\ e.\ particular rearrangements of the tensor values to form a matrix) \cite{kolda} and the usage of matrix norms on these matricizations will lead to interesting and more powerful results. The matrix norms we will be dealing with are the Frobenius or Hilbert-Schmidt norm (which is the standard tensor norm on a matrix), the trace norm and the Ky Fan $k$ norms \cite{HJ}. That is, let $A\in\mathbb{R}^{m\times n}$, then
\begin{align}
||A||&=\sqrt{\sum_{ij}A_{ij}^2}=\sqrt{\sum_i\sigma_i^2},\nonumber\\
||A||_{tr}&=\tr\sqrt{A^TA}=\sum_i\sigma_i,\nonumber\\
||A||_k&=\sum_{i=1}^k\sigma_i,
\end{align}
where $\{\sigma_i\}$ $(i=1,\ldots,\min(m,n))$ denote the singular values of the matrix, which are arranged, as usual, in non-increasing order. Notice that the last Ky Fan norm is the trace norm, i.\ e.\ $||\cdot||_{\min(m,n)}=||\cdot||_{tr}$.

We will define matricizations in the following way: (non) underlined indices are joined together in lexicographical order to give rise to the row (column) indices. Let $A$ be the subset of underlined indices, then we will call that a $A,\bar{A}$ matricization. For example, let $i_j=1,\ldots,n_j$ $\forall j$, then
\begin{equation}
T_{\underline{i_1}i_2\underline{i_3}i_4}=\left(
                                     \begin{array}{cccc}
                                       T_{111k} & T_{121k} & \cdots & T_{1n_21k} \\
                                       T_{112k} & T_{122k} & \cdots & T_{1n_22k} \\
                                       \vdots & \cdot & \cdot & \vdots \\
                                       T_{11n_3k} & \cdots & \cdots & \cdot \\
                                       T_{211k} & \cdots & \cdots & \cdot \\
                                       \vdots & \cdot & \cdot & \vdots \\
                                       T_{21n_3k} & \cdots & \cdots & \cdot \\
                                       \vdots & \cdot & \cdot & \vdots \\
                                       T_{n_111k} & \cdots & \cdots & \cdot \\
                                       \vdots & \cdot & \cdot & \vdots \\
                                       T_{n_11n_3k} & \cdots & \cdots & T_{n_1n_2n_3k} \\
                                     \end{array}
                                   \right),
\end{equation}
where $T_{xyzk}=(T_{xyz1} \cdots T_{xyzn_4})$ is a row vector ($k=1,2,\ldots,n_4$), is a 13,24 matricization. In Dirac notation we would have
\begin{equation}
T_{\underline{i_1}i_2\underline{i_3}i_4}=\sum_{i_1\cdots i_4}T_{i_1i_2i_3i_4}|i_1i_3\rangle\langle i_2i_4|.
\end{equation}
This way of matricizing is a generalization of the concept of matrix unfolding or mode-$n$ matricization which is often used in multilinear algebra \cite{kolda}, which corresponds to matricizations of one index giving rise to the row column and the rest to the column vectors. The matricizations we have defined are quite convenient for the problem at hand because they have a well defined structure under outer product of tensors. Of course, if all the indices of a tensor are joined together the tensor is vectorized, $T_{\underline{i_1i_2i_3}}=vec(T)$, while it is straightforward to check that $T_{\underline{i_1}i_2}W_{{\underline{i_3}i_4}}=(T_{i_1i_2})\otimes(W_{i_3i_4})$. Concatenating these rules a matrix form for any matricization of more involved outer products of tensors can be readily found. For instance,
\begin{align}
T&_{i_{1}\underline{i_{2}}i_3}R_{i_4}S_{\underline{i_{5}i_{6}}}W_{i_7\underline{i_{8}}}\nonumber\\&=(T_{i_2,i_1i_3})\otimes(R_{i_4})^T\otimes vec(S_{i_5i_6})\otimes(W_{i_7i_8})^T.
\end{align}

This is the kind of structure we need because the norms we are going to use are either multiplicative ($||\cdot||$, $||\cdot||_{tr}$ and $||\cdot||_1$) or submultiplicative ($||\cdot||_k$) under tensor products \footnote{This is a consequence of the fact that if $\{\sigma_i\}$ and $\{\sigma'_j\}$ are respectively the singular values of the matrices $A$ and $B$, then the singular values of $A\otimes B$ are $\{\sigma_i\sigma'_j\}$.}. So, analogously to the previous section, we just need to upper bound these quantities for meaningful correlation tensors to obtain conditions for genuine multipartite entanglement. For the sake of simplicity we will consider multiqubit systems in the following subsections.

\subsection{Three qubits}

According to the above discussion the only thing left to be able to derive genuine multipartite entanglement conditions is to obtain upper bounds for the matrix norms of the correlation tensors similarly as we did with the standard norm in the previous section. We have the following

\begin{lemma}\label{2corr}
The 2-body correlation tensor of two qubits satisfies
\begin{equation}\label{knorm2corr}
||T^{(j,l)}_{\underline{i_j}i_l}||_{k}\leq k\quad\forall k,
\end{equation}
with equality iff the two qubits are in a maximally entangled state.
\end{lemma}
\textit{Proof.} We will use the local unitary invariance of the norms of any matricization of the correlation tensors (see Sec.\ \ref{props} below). Notice then that, due to $SO(3)\simeq SU(2)$, the 2-body correlation tensor (i.\ e.\ correlation matrix) can be brought into diagonal form by choosing properly local unitaries in the two subsystems (see e.\ g.\ \cite{hor96}). Since the entries of this matrix are expectation values of observables with eigenvalues 1 or $-1$ we have that $|T_{ii}|\leq1$ $\forall i$. It can be readily checked that this bound is attained by the maximally entangled state (and only by the maximally entangled state because this value of the trace norm of the correlation matrix implies the maximal possible amount of entanglement, as measured, for instance, by the concurrence \cite{yopra}). \hfill$\square$

\begin{lemma}\label{lema3q}
If a pure 3-qubit state is biseparable, then it holds that:
\begin{itemize}
\item[(i)] If the state is fully separable
\begin{equation}
||T_{\underline{i_j}i_li_m}||_{k}\leq 1\quad\forall k.
\end{equation}
\item[(ii)] If the state contains no entanglement across $j|lm$
\begin{equation}\label{noe}
||T_{\underline{i_j}i_li_m}||_{k}\leq \sqrt{3}\quad\forall k.
\end{equation}
\item[(iii)] If the state contains some entanglement across $j|lm$
\begin{equation}\label{somee}
||T_{\underline{i_j}i_li_m}||_{k}\leq k\quad\forall k.
\end{equation}
\end{itemize}
\end{lemma}
\textit{Proof.} We will use repeatedly the upper bounds (\ref{snorm1corr}), (\ref{snorm2corr}) and (\ref{knorm2corr}).
\begin{itemize}
\item[(i)]
\begin{align}
||T_{\underline{i_j}i_li_m}||_{k}&=||T^{(j)}_{\underline{i_j}}T^{(l)}_{i_l}T^{(m)}_{i_m}||_k=||(T^{(j)})\cdot(T^{(l)}\otimes T^{(m)})^T||_k\nonumber\\
&=||T^{(j)}||||T^{(l)}\otimes T^{(m)}||\nonumber\\
&=||T^{(j)}||||T^{(l)}|| ||T^{(m)}||=1.
\end{align}
\item[(ii)]
\begin{align}
||T_{\underline{i_j}i_li_m}||_{k}&=||T^{(j)}_{\underline{i_j}}T^{(l,m)}_{i_li_m}||_k=||(T^{(j)})\cdot vec(T^{(l,m)})^T||_k\nonumber\\
&=||T^{(j)}|| ||T^{(l,m)}||\leq\sqrt{3}.
\end{align}
\item[(iii)]
\begin{align}
||T_{\underline{i_j}i_li_m}||_{k}&=||T^{(j,l)}_{\underline{i_j}i_l}T^{(m)}_{i_m}||_k=||(T^{(j,l)})\otimes (T^{(m)})^T||_k\nonumber\\
&\leq||T^{(j,l)}||_k ||T^{(m)}||\leq k.
\end{align}
\end{itemize}
\hfill$\square$

From Lemma \ref{lema3q} we read 3 sufficient conditions for genuine multipartite entanglement, namely that the norm of any of the three possible matricizations of the full correlation tensor is greater than $\sqrt{3}$, $2$ and $3$ for $||\cdot||_1$, $||\cdot||_2$ and $||\cdot||_{tr}$ respectively. To illustrate the power of these conditions consider that the singular values of these matricizations are $\{1.414,1.414,0\}$ for the GHZ state and $\{1.374,0.943,0.943\}$ for the W state. Hence, the last two conditions can detect genuine multipartite entanglement. Notice that these states are symmetric, so all matricizations of the full correlation tensor are equal; however, for general states our ability to detect a state as genuinely multipartite entangled might depend on the choice of matricization. To avoid this and to obtain a stronger condition which takes into account a combination of the bounds of Lemma \ref{lema3q} rather than just picking one of them we introduce the average matricization norm $||M(T_{i_1i_2i_3})||=(||T_{\underline{i_1}i_2i_3}||+||T_{i_1\underline{i_2}i_3}||+||T_{i_1i_2\underline{i_3}}||)/3$, which leads to
\begin{theorem}\label{th3q}
If for a 3-qubit state it holds that
%\begin{equation}
%||M(T_{i_1i_2i_3})||_2>\frac{4+\sqrt{3}}{3}\simeq 1.91
%\end{equation}
%or
\begin{equation}
||M(T_{i_1i_2i_3})||_{k}>\frac{2k+\sqrt{3}}{3},%\simeq 2.58,
\end{equation}
then the state contains genuine multipartite entanglement.
\end{theorem}
\textit{Proof.} Simply use that any biseparable state can be written as $\rho_{bs}=\sum_kp_k\rho_{12}^k\otimes\rho_3^k+q_k\rho_{13}^k\otimes\rho_2^k+r_k\rho_{23}^k\otimes\rho_1^k$ and combine properly the bounds (\ref{noe}) and (\ref{somee}). \hfill$\square$

Thus, Theorem \ref{th3q} allows to detect genuine multipartite entanglement in mixtures of the GHZ and W states with white noise for noise levels of $p_{GHZ}\lesssim0.324$ and $p_{W}\lesssim0.209$. Notice that it is known that these states are genuinely multipartite entangled iff $p_{GHZ}\lesssim0.571$ and $p_{W}\lesssim0.521$ \cite{Guehnetaming}.

\subsection{Four qubits}

Now, similarly to previous sections, we need upper bounds to the norms of the matricizations of the 3-body correlation tensor. However, it is not clear which states should attain the maximum values of these norms in opposition to the 2-body case, where the maximally entangled state, as intuition would suggest, does the job. Moreover, numerics indicate that $\max_{|\psi\rangle}||T_{\underline{i_j}i_li_m}||_{tr}\simeq3.272$ for a state which, although close to the W state, has no simple mathematical structure. This indicates that devising a systematic procedure to find the maximum value of these norms similarly as we did with the standard norm in Sec.\ \ref{standard} might be very hard. Nevertheless, it turns out that we can use this procedure to obtain reasonable estimates by using the equivalence of the norms: $||\cdot||_{k}\leq\sqrt{k}||\cdot||$.

\begin{lemma}\label{3corr}
The 3-body correlation tensor of three qubits satisfies $||T_{\underline{i_j}i_li_m}^{(j,l,m)}||_{k}\leq2\sqrt{k}$ $\forall k$.
\end{lemma}
\textit{Proof.} The result follows from $||T_{\underline{i_j}i_li_m}^{(j,l,m)}||\leq2$. To see this we proceed as in Sec.\ \ref{standard}. The fact that $\tr\rho_{jlm}^2=1$ translates to
\begin{equation}
\sum_{s=j,l,m}||T^{(s)}||^2+\sum_{s<q}||T^{(s,q)}||^2+||T^{(j,l,m)}||^2=7.
\end{equation}
The minimum possible values of the norms of the lowest order correlation tensors are $||T^{(s)}||=0$ and $||T^{(s,q)}||=1$ $\forall s,q$ (since the 1-qubit reduced density matrices can be maximally mixed but the highest mixing allowed by the 2-qubit reduced density matrices is them being equal to the identity in a two-dimensional subspace). Therefore, $||T^{(j,l,m)}||\leq2$, which is attained by the GHZ state. Finally, notice that the Hilbert-Schmidt norm of any matricization of a tensor equals its standard norm as tensor.\phantom{xxxxx}\hfill$\square$

Notice that Lemma \ref{3corr} provides accurate estimates as the trace norm bound $2\sqrt{3}\simeq3.464$ is quite close to the numerical maximum given above while the Ky Fan 2 norm bound is actually sharp since it is attained by the GHZ state.

Now, we can proceed as in Lemma \ref{lema3q} to upper bound $||T_{\underline{i_ji_l}i_mi_s}||_k$. Then, for pure biseparable states, one obtains the bounds $2\sqrt{k}$ (for $k\leq3$ and $2\sqrt{3}$ otherwise) if $T_{i_ji_li_mi_s}=T^{(j)}_{i_j}T^{(l,m,s)}_{i_li_mi_s}$ or $T_{i_ji_li_mi_s}=T^{(j,l,m)}_{i_ji_li_m}T^{(s)}_{i_s}$ (i.\ e.\ the state is biseparable in one subsystem versus the other three and we consider the two possibilities that the two indices giving rise to the row of the matricization either belong to unentangled or entangled particles), 3 if $T_{i_ji_li_mi_s}=T^{(j,l)}_{i_ji_l}T^{(m,s)}_{i_mi_s}$ and $k$ if $T_{i_ji_li_mi_s}=T^{(j,m)}_{i_ji_m}T^{(l,s)}_{i_li_s}$. One could also consider 1 vs. 3 matricizations of the full correlation tensor; however, one finds that $||T_{\underline{i_j}i_li_mi_s}||_k\leq\sqrt{3}k$ for $T_{i_ji_li_mi_s}=T^{(j,l)}_{i_ji_l}T^{(m,s)}_{i_mi_s}$ (i.\ e.\ any 2 vs. 2 biseparable state), which turns out to be a weak condition and, then, it is better not to take these matricizations into account. Combining all the above bounds, defining the 2 vs.\ 2 average matricization norm $||M_{22}(T_{i_1i_2i_3i_4})||=(||T_{\underline{i_1i_2}i_3i_4}||+||T_{\underline{i_1}i_2\underline{i_3}i_4}||+||T_{\underline{i_1}i_2i_3\underline{i_4}}||)/3$ and proceeding as in Theorem \ref{th3q} we have

\begin{theorem}\label{th4q}
If for a 4-qubit state one of the following inequalities holds
\begin{equation}
||M_{22}(T_{i_1i_2i_3i_4})||_k>\left\{\begin{array}{c}
                                        2\sqrt{k}\quad\quad\,\;\;\, 1\leq k\leq3 \\
                                        1+2k/3\quad 4\leq k\leq9
                                      \end{array}\right.,
\end{equation}
then the state contains genuine multipartite entanglement.
\end{theorem}

With this, genuine multipartite entanglement is detected in the GHZ state with a white noise tolerance of $p_{GHZ}\lesssim0.307$ and for the Dicke states of 1 and 2 excitations we have respectively $p_{D_1}\lesssim0.018$ and $p_{D_2}\lesssim0.328$. From \cite{Guehnetaming} we know that there is genuine multipartite entanglement iff $p_{GHZ}\lesssim0.533$ and if $p_{D_2}\lesssim0.539$.

\begin{figure}[h!]
\begin{center}

\includegraphics[width=7cm, keepaspectratio=true]{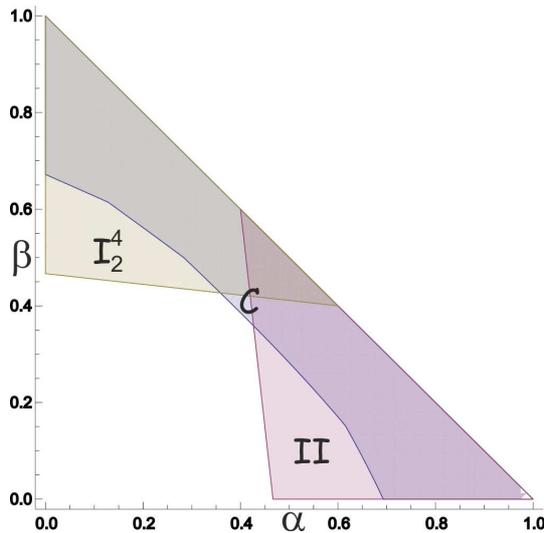}
\caption{(Color online) Here the parameter regions for which the state $\rho=\alpha \rho_{GHZ(2)}+\beta \rho_{D(2)}+\frac{1-\alpha-\beta}{16}\id$ exhibits genuine multipartite entanglement are identified. The GHZ and two-excitation Dicke state for four qubit systems are defined as $\rho_{GHZ(2)}=|GHZ(2)\rangle\langle GHZ(2)|$ with $|GHZ(2)\rangle:=\frac{1}{\sqrt{2}}\sum_{i=0}^{1}|i,i,i\rangle$ and $\rho_{D(2)}=|D_2^4\rangle\langle D_2^4|$ with $|D_2^4\rangle=\frac{1}{\sqrt{6}}(|0011\rangle+|0101\rangle+|1001\rangle+|1010\rangle+|1100\rangle+|0110\rangle)$. The (red) region labeled II, uses criterion II from Ref.~\cite{HMGH1} optimized numerically over all local unitary representations of the density matrix. The (yellow) region labeled $I_2^4$, uses criterion $I_2^4$ from Ref.~\cite{HESGH1} optimized numerically over all local unitary representations of the density matrix. The numerical optimization was performed using the composite parametrization from Ref.~\cite{compp}. The (blue) region labeled C, shows the region detected to be genuinely multipartite entangled using theorem \ref{th4q}.}\label{example3}
\end{center}
\end{figure}

These examples indicate that the matricization approach is more powerful than that of the standard norm and that it can detect different classes of entangled states. Interestingly, the detection capability of Theorems \ref{th3q} and \ref{th4q} is already comparable to \cite{HMGH1,HESGH1} for qubits as shown in Figure \ref{example3}.

Using the matricization approach we have thus constructed versatile criteria, detecting genuine multipartite entanglement in a broad variety of cases. All famous examples of four qubit multipartite entangled states are detected (GHZ-, W-, Dicke- and Singlet-state) using the same criterion, without any optimization involved as the norms of any matricization of a correlation tensor is invariant under local unitary transformations on the density matrix (see Sec.\ \ref{props}). Although for some specific states optimizing over all possible witnesses can yield a higher noise resistance in some cases, a comparable result is achieved in a computationally far more efficient way. As shown in Figure \ref{example3}, there even exist states that were not detected to be genuinely multipartite entangled with any of the optimized criteria so far.
\section{Detection of non-fully separable states}\label{full}

As mentioned above, we can also use correlation tensors to discriminate states containing some form of entanglement and fully separable states. This has already been carried out in \cite{mbloch},
where the authors show that that for fully separable states an upper bound on the trace norm of the full correlation tensor must hold for any matricization of the form one particle versus the
rest (i.\ e.\ matrix unfoldings). However, remarkably, our picture allows not only for a very simple proof of this fact, but, also, for a significantly stronger result since we can show that
such a bound must hold for \textit{any possible} matricization of the correlation tensor.

\begin{theorem}\label{fullsep}
For any fully separable state any matricization of the full correlation tensor must fulfill
\begin{equation}
||T_{\underline{i_1\cdots i_k}i_{k+1}\cdots i_n}||_{tr}\leq\prod_{j=1}^n\sqrt{\frac{2(d_j-1)}{d_j}},
\end{equation}
i.\ e.\ $k=1,\ldots,n-1$ and all possible permutations of the particles are taken into account.
\end{theorem}
\textit{Proof.} Since the correlation tensor of a fully separable pure state must fully factorize into the 1-body correlation tensors it is straightforward to note that
\begin{align}
||&T_{\underline{i_1\cdots i_k}i_{k+1}\cdots i_n}||_{tr}=||T^{(1)}_{\underline{i_1}}\cdots T^{(k)}_{\underline{i_k}}T^{(k+1)}_{i_{k+1}}\cdots T^{(n)}_{i_n}||_{tr}\nonumber\\
&=||(T^{(1)}\otimes\cdots\otimes T^{(k)})\cdot(T^{(k+1)}\otimes\cdots\otimes T^{(n)})^T||_{tr}\nonumber\\
&=||T^{(1)}\otimes\cdots\otimes T^{(k)}||\,||T^{(k+1)}\otimes\cdots\otimes T^{(n)}||\nonumber\\
&=\prod_j||T^{(j)}||
\end{align}
Using Eq.\ (\ref{snorm1corr}) the proof is finished.\hfill$\square$

Notice that similar bounds hold as well for lower order correlation tensors. Of course, one could also consider other matrix norms but the trace norm yields the most powerful condition as they all lead to the same bound. In particular, if one considers the standard norm one obtains the criterion of \cite{john}, which is then proved to be strictly weaker than our Theorem \ref{fullsep} as $||\cdot||_{tr}\geq||\cdot||$.

Theorem \ref{fullsep} is then clearly stronger than the criterion of \cite{mbloch} since it contains matrix unfoldings as a particular case while the other matricizations can further
restrict the set of fully separable states leading to a substantially more powerful detection of entangled states. For instance, consider the Dicke states of four qubits for which the matrix
unfolding (i.\ e.\ the criterion of \cite{mbloch}) detects entanglement in some form up to white noise levels of 0.698 (1 excitation) and 0.807 (2 excitations), while the 12,34 matricization
rises these levels to 0.732 and 0.842 respectively. As we will see in more detail with the examples below this seems to be a general feature.

A very interesting conclusion of our study of multipartite entanglement detection with correlation tensors is the fact that the very same piece of information can be used to decide both non
full separability and genuine multipartite entanglement. This gives Theorem \ref{fullsep} an advantage over other conditions for the detection of some form of entanglement in multipartite
states, since used together with Theorems \ref{th3q} and \ref{th4q} one can furthermore discriminate when genuine multipartite entanglement is present. Moreover, as discussed in the introduction, the correlation
tensors take into account all $m$-body correlations while many important conditions rely only on two-point correlations. This
constrains the power of these conditions since, for instance, they cannot detect the important class of graph states as it has been shown in \cite{git}. In particular, this means that,
contrary to our case, they cannot detect the GHZ state, one of the most paradigmatic multipartite entangled states. As we will see with some examples this limitation is extendable to other
classes in which interactions of more than two particles are somehow relevant (e.\ g.\ graph states cannot be nondegenerate ground states of Hamiltonians containing at most two body
interactions \cite{graphhamiltonians}). On the contrary, Theorem \ref{fullsep} turns out to be quite efficient in these cases.

\begin{figure}[t]
\includegraphics[scale=0.5]{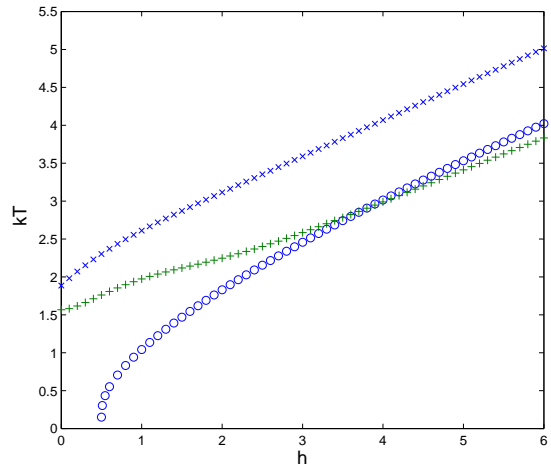}
\caption{ (Color online) Maximal temperature for which entanglement is detected in the thermal states of $H_1$ by: Theorem \ref{fullsep}, i.\ e.\ any matricization of the full correlation tensor, (x marks), the criterion of \cite{mbloch}, i.\ e.\
matrix unfoldings of the full correlation tensor (pluses) and OSSI (circles).}\label{thermal1}
\end{figure}

\begin{figure}[t]
\includegraphics[scale=0.5]{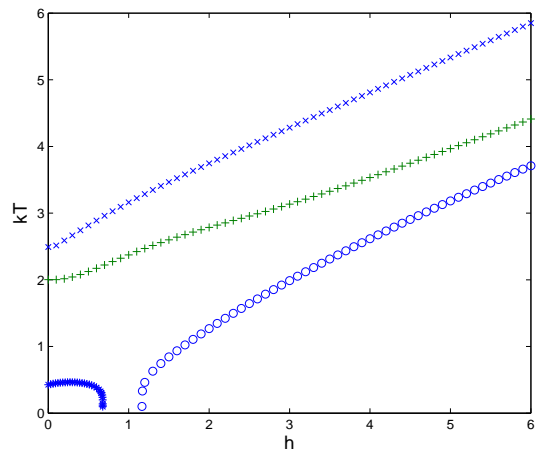}
\caption{ (Color online) The same as in Figure \ref{thermal1} for $H_2$. The area in the down-left corner represents genuine multipartite entanglement as detected by Theorem \ref{th4q}.}\label{thermal2}
\end{figure}

Let us start by considering the 4-qubit 3-body interaction
Hamiltonian with transversal magnetic field of strength $h$
\begin{equation}
H_1=\sum_j(-\sigma_z^{(j-1)}\sigma_x^{(j)}\sigma_z^{(j+1)}+h\sigma_x^{(j)}),
\end{equation}
where periodic boundary conditions are assumed and the superscript indicates on which qubit the operation is acting. For brevity, the tensor products and the identity operation are omitted. In Figure \ref{thermal1} we plot the detection efficiency of correlation tensors and the optimal spin squeezing inequalities (OSSI) \cite{ossi} for the thermal states of $H_1$ \footnote{The OSSI is the
strongest entanglement criterion possible if only averaged two-particle correlations are used. Hence, in what follows we will take it as good indicator of the power of all the aforementioned
criteria based on two-body correlations.}, i.\ e.\
\begin{equation}
\rho(kT,h)=\frac{\exp(-H_1/kT)}{\tr(\exp(-H_1/kT))}.
\end{equation}
The superiority of our condition is not so surprising as it is known that the (nondegenerate) ground state of $H_1$ when $h=0$ is the cluster state, a graph state. Moreover, one can consider a
slight variation of $H_1$, namely
\begin{equation}
H_2=\sum_j[-\sigma_z^{(j-1)}(\sigma_x^{(j)}+\sigma_y^{(j)}+\sigma_z^{(j)})\sigma_z^{(j+1)}+h\sigma_x^{(j)}],
\end{equation}
for which the thermal ground state when $h=0$ is not a graph state \footnote{This is because there is a finite set of LU inequivalent 4-qubit graph states, which
can be seen not to be the thermal ground state in this case.}. Nevertheless, correlation tensors remain more powerful than the OSSI as shown in Figure \ref{thermal2}, and, furthermore, a region of genuine
multipartite entanglement can be identified with them. Let us mention that these Hamiltonians are not artificial, they can arise as effective interactions in the context of
optical lattices in a triangular configuration \cite{optlat} and their statistical mechanical properties are a subject of current research \cite{clusterising}.

In the same vein we have considered the class of locally maximally entanglable (LME) states \cite{lme}, which generalize graph states. These states can be generated with nonlocal gates acting on a
product state, these operations being generalized phase gates with $m$-body interactions. We have checked Theorem \ref{fullsep} in randomly generated 4-qubit LME states and they were always
found to be entangled \footnote{The states are detected both with a matrix unfolding and 2 vs.\ 2 matricization, but the latter provides a larger violation, which, for instance, translates
in a stronger white noise tolerance}. Moreover, in most cases they were found to be genuinely multipartite entangled by Theorem \ref{th4q}. On the contrary, the OSSI always
failed for these states \footnote{Notwithstanding, some non-generic LME states (i.\ e.\ of measure zero) can be found to be detected by OSSI. Nevertheless, we found a
greater white noise tolerance for them using Theorem \ref{fullsep}.}.

Last, we have checked that Theorem \ref{fullsep} is able to detect entangled states with a positive partial transposition (PPT). This was already known in the bipartite case \cite{bloch,yojpa}. However, one may wonder if this is still possible in the multipartite case for states which are PPT with respect to every possible bipartition. This question is answered in the affirmative. We have considered the 3--qubit Hyllus state \cite{hyllus} which depends on one free parameter. For all values of this parameter, despite not fully separable, this state is not only PPT with respect to every possible bipartition but separable with respect to every possible bipartition. It can be readily checked that Theorem \ref{fullsep} detects this state as entangled for many values of the free parameter. This shows the strength of our condition since the entanglement in this state is very weak: it is fully bound entangled, i.\ e.\ the state cannot be distilled even if different parties act together.

\section{Mathematical properties and experimental implementation}\label{props}

As we have shown the correlation tensor provides a powerful tool, enabling the detection of a wide range of multipartite entangled states. It is worth pointing out that
\begin{fact}
All the quantities needed to apply the conditions presented here (Theorems \ref{thsnorm}--\ref{fullsep}) are invariant under local unitary (LU) transformations on the density matrix.
\end{fact}

More precisely, all the norms considered here ($||\cdot||$, $||\cdot||_k$ and $||\cdot||_{tr}$) of any matricization of any correlation tensor are invariant under these transformations. This is because of the well-known fact that LU operations acting on the density matrix correspond to rotations in correlation space (see e.\ g.\ \cite{schmah,bloch,john}). Hence, the transformation $\rho\rightarrow U_1\otimes\cdots\otimes U_n\rho U_1^\dag\otimes\cdots\otimes U_n^\dag$ amounts to $T_{\underline{i_1\cdots i_k}i_{k+1}\cdots i_n}\rightarrow(O_1\otimes\cdots\otimes O_k)T_{\underline{i_1\cdots i_k}i_{k+1}\cdots i_n}(O_{k+1}\otimes\cdots\otimes O_n)$ for any matricization of a correlation tensor $T$ for some real orthogonal matrices $\{O_i\}$. The result then follows because the aforementioned norms are all unitarily invariant \cite{HJ}.

This is a convenient property which is not shared by other conditions for genuine multipartite entanglement \cite{guehnecrit,HMGH1}. Although one can nevertheless take the effort of optimizing over LUs for the application of these criteria in practice, LU invariance is a very satisfactory property from the theoretical point of view since this is a fundamental property of entanglement. Moreover, this could lead to the use of these norms of correlation tensors not only as qualitative indicators of entanglement but also as quantitative tools (this is already the case for the trace norm in the bipartite case \cite{yopra} and for the standard norm in the multipartite case for non-full separability \cite{yojpa,blochmes}). Furthermore, LU invariance is also quite convenient from the experimental point of view since this implies that the local measurement settings of each party need not be aligned with the others \footnote{Besides their fundamental limitations for entanglement detection, Bell-like inequalities can be used to detect genuine multipartite entanglement in a device independent way (i.\ e.\ with no need to trust the measurement devices) \cite{diew}. This can be quite convenient from the experimental point of view since, as this reference argues, the presence of tilts in the measurements reduces notoriously the entanglement detection capability of entanglement witnesses. Our scheme is of course not device independent but measurements only need to be locally perfect (each party has to measure othogonal observables but they do not need to match those of other parties), which can be very useful when the different parties are far away.}.

Regarding experimental implementation, it is also worth discussing the number of measurements required to use our criteria. Although knowledge of the full correlation tensor requires less parameters than that of the density matrix, the statistical data that needs to be collected to reconstruct it allows, however, to do state tomography. Nevertheless, partial knowledge of a correlation tensor allows as well to implement our conditions as the norm of a matrix can be estimated from below by knowing just some of its entries. This is clear for the standard norm from Eq.\ (\ref{snorm}). Using again the equivalence of the norms, e.\ g.\ $||\cdot||_{tr}\geq||\cdot||$, one can lower bound the other norms. In addition to this, if particular entries of a matrix are known one can directly obtain lower bounds for the Ky Fan norms. For instance, we have that
\begin{equation}
||A||_{tr}\geq\sum_i|a_{ii}|,
\end{equation}
and, also that the Ky Fan norm of any principal submatrix is a lower bound for the Ky Fan norm of the full matrix. Furthermore, the minimum value of the trace norm of a matrix subject to the knowledge of some of its entries can be efficiently computed using convex optimization techniques (see e.\ g.\ \cite{gross}). Hence, if the numbers given by these lower bounds are already large enough to violate the inequalities given by Theorems \ref{thsnorm}--\ref{fullsep}, one can then conclude with certainty the presence of genuine multipartite entanglement or non full separability with considerably fewer measurements.

Genuine multipartite entanglement is usually addressed in terms of entanglement witnesses. They are locally measurable and the number of required measurements can scale very favorably with the system size (i.e. polynomially in Refs.~\cite{HMGH1,HESGH1} and even linearly in Ref.~\cite{HSGSBH1}). However, this limited number of measurements of course also severely limits the number of states that are detected by such criteria (e.g. GHZ states using the criteria from Refs.~\cite{HMGH1,HESGH1}). Nevertheless, if one has theoretical expectations of what the state should look like, one can then use a suitable criterion which should be able to detect these states.
Our presented framework can allow as well for a versatile detection of any state that is detected by our criteria using only a very limited number of measurements as in many cases a considerable number of expectation values (i.\ e.\ elements of the correlation tensors) are zero.
\begin{itemize}
\item First calculate all correlation tensor elements of the theoretically expected state in an experiment
\item Second, only perform the measurements corresponding to exactly these elements
\item Then lower bound the norm of the correlation tensor using only this limited amount of elements
\end{itemize}
If this lower bound exceeds the threshold of any of our inequalities it is certain that it contains entanglement. This implies that in cases where entanglement witnesses are applicable (i.e. some prior expectations of the state), we can apply our criteria in a just as experimentally feasible way. E.g. for three qubit GHZ states the criterion from Ref.~\cite{HMGH1} requires seven local measurement settings which is exactly the number of correlation tensor elements that have to be ascertained.

\section{Conclusions}\label{conclusions}

We have provided a general framework to detect different classes of multipartite entanglement in systems of arbitrary dimension using as a main tool correlation tensors. In particular, considering several norms on these objects, we have shown that different upper bounds can be established such that violations signal the presence of either non-full separability or genuine multipartite entanglement. Regarding genuine multipartite entanglement, we have explicitly worked out the case of tripartite qudit (Theorems 1 and 2) and four partite qubit (Theorem 3) systems. The approach, however, can be generalized to an arbitrary number of subsystems and dimensions in a tedious but systematical way. It would be interesting to study in the future if this procedure can be rendered more straightforward. This is the case for our sufficient condition for non-full separability (Theorem 4), which has a very simple proof for arbitrary number of subsystems and dimensions.

We have demonstrated as well with exemplary cases that our approach can improve and complement previous comparable criteria in both the non-full separability and genuine multipartite entanglement scenarios. Furthermore, the entries of the correlation tensors are directly related to measurable quantities and we have discussed how to estimate the relevant norms with fewer measurements to ease the experimental implementation. Last, our norms are all LU invariant, which besides some implementation advantages, is a satisfactory property from the theoretical point of view. This suggests that they might be connected to the quantification of entanglement with entanglement measures, thus not only providing qualitative information. This would be particularly interesting in the genuine multipartite entanglement case, where the first quantification steps have been taken in \cite{Guehnetaming,crazychin}. We leave for future research the question of whether our norms could provide a (rough) easily computable quantification of genuine multipartite entanglement.

Another interesting point for future research is the connection between the norms of the correlation tensor and non-locality. It has been shown in \cite{horchsh} that the set of two-qubit states violating the CHSH inequality is characterized in terms of the singular values of the correlation matrix, i.\ e.\ a state violates CHSH iff the sum of the squares of the two largest singular values of $T^{(1,2)}$ is greater than 1. It would be interesting to study if this connection can be extended to the multipartite regime using the norms and correlation tensors we have introduced here and to establish analogous characterizations for other Bell inequalities.

\section*{ACKNOWLEDGMENTS}

We thank B.C. Hiesmayr, B. Jungnitsch, B. Kraus, T. Moroder, T. Paterek and Ch.\ Spengler for useful discussions. J.I. de V. acknowledges financial support from the Austrian Science Fund (FWF): Y535-N16 and
F40-FoQus F4011-N16 and M.H. support from the science fund FWF-P21947N16.

\end{document}